\documentclass{elsart}
\usepackage{graphicx}
\usepackage{amsmath, amssymb}

\newcommand{\Ha}{\mbox{$\cal H$}}

\begin{document}

\begin{frontmatter}

\title{Free-Surface Hydrodynamics in the conformal variables}
\author[NGU,Tucson,Landau,Phian]{V.E. Zakharov}
\author[NGU,Landau]{A.I. Dyachenko,\corauthref{cor}}
\ead{alexd@landau.ac.ru}
\corauth[cor]{Corresponding author.}

{
\ead{zakharov@math.arizona.edu}
}
\address[NGU]{Novosibirsk State University, Pirogova 2,Novosibirsk-90, 630090, Russia}
\address[Tucson]{Department of Mathematics, University of
Arizona, Tucson, AZ, 857201, USA}
\address[Phian]{Physical Institute of RAS,
Leninskiy prospekt, 53, Moscow, 119991, Russia}
\address[Landau]{Landau Institute for Theoretical Physics,
2 Kosygin str., Moscow, 119334, Russia}

\begin{abstract}
The potential flow of two-dimensional ideal incompressible fluid
with a free surface is studied.
Using the theory of conformal mappings and Hamiltonian formalism
allows us to derive exact equations of surface evolution. Simple form of the
equations helped to discover new integrals of motion. These integrals
 are connected with
the analytical properties of conformal mapping and complex velocity.
Simple form of the equations also makes the numerical simulations
of the free surface evolution very straightforward.

In the limit of almost flat surface the equations can be reduced to
the Hopf equation.
\end{abstract}


\end{frontmatter}

\section{Basic equations}
We study the potential flow of two-dimensional ideal incompressible fluid.
The fluid occupies a half-infinite domain
\begin{equation}
-\infty < y < \eta(x,t), \,\,\,\,\,\, -\infty < x < \infty.
\end{equation}
The flow is potential, so that:
\begin{equation}
v = \nabla \Phi, \,\,\,\,\,\,\,\,\,\,\,\,\,\,\,\,\,\,\,\,\,\,\,
\Phi\vert_{y=\eta(x,t)} = \psi(x,t).
\end{equation}

Boundary conditions on the surface are standard:
\begin{eqnarray}\label{Eq12}
\frac{\partial\Phi}{\partial t} &+& \frac{1}{2}|\nabla \Phi|^2 +g\eta = P, \cr
\frac{\partial\eta}{\partial t} &+& \eta_x \Phi_x = \Phi_y
\end{eqnarray}
at $y = \eta(x,t)$ and
\begin{eqnarray}\label{Eq22}
\frac{\partial \Phi}{\partial y} = 0,  y \rightarrow -\infty, \cr
\frac{\partial \Phi}{\partial x} = 0,  |x| \rightarrow \infty.
\end{eqnarray}
Here $g$ is the gravity acceleration and $P$ - constant pressure at the
surface (Let $P=0$).
It is known
(
\cite{Z68}), that the shape of surface $\eta(x,t)$
and the
potential on the surface $\psi(x,t)$ form pair of canonically conjugated
variables obeying the Hamiltonian equations:
\begin{equation}\label{pair}
\frac{\partial \eta}{\partial t} = \frac{\delta {\cal H}}{\delta \psi}, \qquad
\frac{\partial \psi}{\partial t} = -\frac{\delta{ \cal H}}{\delta \eta}.
\end{equation}
Here ${\cal H}$ is Hamiltonian function (total energy of the fluid):
$${\cal H} = T = \frac{1}{2}\int_{-\infty}^{\infty}dx\int_{-\infty}^{\eta(x,t)}
|\nabla\Phi|^2 dy$$
kinetic energy of the fluid.
Along with the energy there are three more integrals of motion,
amount of fluid:
\begin{equation}\label{mass}
\frac{\partial}{\partial t}\int_{-\infty}^{\infty}\eta(x,t)dx = 0,
\end{equation}
and vertical and horizontal momenta:
\begin{eqnarray}\label{momenta}
\frac{\partial}{\partial t}\int_{-\infty}^{\infty}dx
\int_{-\infty}^{\eta(x,t)} \phi_y dy &=& 0 , \cr
\frac{\partial}{\partial t}\int_{-\infty}^{\infty}dx
\int_{-\infty}^{\eta(x,t)} \phi_x dy &=& 0.
\end{eqnarray}

The equations (\ref{Eq12}) are functionally nonlinear and
can be hardly studied. The most important known solutions were
derived by Dirichlet in 1860. In these solutions a shape of the surface
is a quadric (ellipse, hyperbola and parabola). Dirichlet solutions
are described in details in \cite{L-H76}.

Equations (\ref{pair}) minimize the action
\begin{equation}\label{action}
S = \int L dt
\end{equation}
\begin{equation}\label{Lagrangian}
L = \int_{-\infty}^{\infty} \psi\eta_t dx -{\cal H}.
\end{equation}

Starting from this point let us forget for a while about hydrodynamics, and
consider more general case. Namely,
let's think of $\cal H$ as some arbitrary functional of $\psi$ and $\eta$.
Then, let's introduce instead of $\eta(x,t)$ new complex variable $z(u,t)$
in the following way: $z(w,t)$ is complex function of complex variable
$w=u + iv$.
It is analytic in the lower half-plane of complex variable $w$. Its real and
imaginary parts, given on the real axes ($z(u,t) = x(u,t) +iy(u,t)$)
parametrically define $y(x,t) = \eta(x,t)$.

Another words, $z(w,t)$ is the conformal mapping of the domain,
bounded by the curve $\eta(x,t)$
to the lower half-plane of $w$,
$$w = u+ iv, \qquad -\infty <u <\infty, \quad -\infty <v <0.$$

In the new variables Lagrangian takes the form
\begin{eqnarray}
L = \int_{-\infty}^{\infty}\psi[y_t x_u-x_t y_u]du
-\int_{-\infty}^{\infty}f(y-\hat H(x-u))du
-\cal H.\nonumber
\end{eqnarray}
Here $\hat H$ is the Hilbert transformation, that provide the relation
between real and imaginary parts of analytic function:
\begin{equation}\label{Hilbert}
y = \hat H(x-u) = {1\over\pi}P.V.\int_{-\infty}^{\infty}\frac{x(u')-u'}{u'-u}du'.
\end{equation}
Motion equations should be found from the condition $\delta S=0$. When
performing variation one should take into account that $x$ and $y$ are
connected by relation (\ref{Hilbert}), and Hamiltonian function $\cal H$
can be considered as a functional, depending only on $\psi$ and $y$.
Hence, one can put $\frac{\delta\cal H}{\delta x} = 0$.

The condition $\delta S = 0$ leads to the following ``implicit''
equations of motion
\begin{equation}\label{1implicit}
y_t x_u-x_t y_u = \frac{\delta\cal H}{\delta\psi}
\end{equation}
\begin{equation}\label{2implicit}
\psi_t x_u - x_t\psi_u +f = -\frac{\delta\cal H}{\delta y}
\end{equation}
\begin{equation}\label{3implicit}
y_u \psi_t - y_t \psi_u -\hat Hf = 0
\end{equation}
From (\ref{3implicit}) one can find
\begin{equation}\label{4implicit}
f = \hat H(y_t \psi_u - y_u \psi_t).
\end{equation}
While $\psi$ satisfies the equation
\begin{eqnarray}\label{2of2}
\psi_t x_u -x_t\psi_u -\hat H(y_u\psi_t-y_t\psi_u) =
-\frac{\delta\cal H}{\delta y }.
\end{eqnarray}
Equations (\ref{1implicit}) and (\ref{2of2}) are resolved with respect
to variational derivatives $\frac{\delta\cal H}{\delta\psi }$ and
$\frac{\delta\cal H}{\delta y }$. They can be written in the explicit
Hamiltonian form:
\begin{eqnarray}\label{omega}
\Omega_{11}\dot y + \Omega_{12}\dot\psi &=& \frac{\delta\cal H}{\delta y },\cr
-\Omega_{12}^+\dot y &=& \frac{\delta\cal H}{\delta\psi}.
\end{eqnarray}
Here the operators comprising the symplectic form $\Omega_{ij}$ are:
\begin{eqnarray}\nonumber
\Omega_{11} &=& -(\psi' \hat H +(\hat H\psi')),\cr
\Omega_{12} &=& -x' +\hat Hy', \hspace{2cm}\Omega_{21}^+ = x' +y'\hat H.
\end{eqnarray}

As far as the equations (\ref{omega}) are obtained directly from the
variational principle, the symplectic form $\Omega_{ij}$ is close and
nondegenerated. It means, in particulary that equations (\ref{omega})
have no Casimirs - the constant of motion which do not depend on a choice
of the Hamiltonian function $\cal H$.

This is remarkable that equations \ref{1implicit}) and (\ref{2of2}) in a general
case can for an arbitrary $\cal H$
be resolved explicitly with respect to time-derivatives.
To do this we introduce projective oprerators
$$\hat P^{\pm} = \frac{1}{2}(1 \pm i\hat H), \hat P^{+}\hat P^{-} = 0,
\hat {P^{\pm}}^2 = \hat P^{\pm}$$.
Any complex-valued function $\phi(u)$, $-\infty < u < \infty$ can be presented as follows
$$\phi(u) = \hat P^{+}\phi(u) + P^{-}\phi(u)$$.
Here $P^{\pm}\phi(u)$ are analytic functions in the upper and lower half-planes.

Mention first that equation (\ref{1implicit}) is equivalent to the equation
\begin{equation}\label{reverse1}
\frac{\delta\cal H}{\delta\psi} = -\frac{i}{2}(\dot z\bar z^\prime -
\dot{\bar z} z^\prime).
\end{equation}

We denote the Jacobian $J$ of the conformal mapping
$$J = |z'|^2.$$
Dividing (\ref{reverse1}) by $J$ and applying projective operator
$$\hat P^{-} =\frac{1}{2}(1 + i\hat H)$$
one can get from (\ref{reverse1}) the following equation:
\begin{eqnarray}\label{1complex}
\dot z &=& iUz^\prime,\cr
U &=& 2 \hat P^-(\frac{1}{J}\frac{\delta\cal H}{\delta\psi})
\end{eqnarray}

For $\dot y$ and $\dot x$ one can get the equations:
\begin{eqnarray}\label{ReCo}
\dot y &=& (x' - y' \hat H)\frac{1}{J}\frac{\delta\cal H}{\delta\psi},\cr
\dot x &=& -(y' + x' \hat H)\frac{1}{J}\frac{\delta\cal H}{\delta\psi}.
\end{eqnarray}
Excluding $\psi_t$ from (\ref{2implicit}) and (\ref{3implicit})
and using equation (\ref{1implicit}) one can
obtain the folowing relation:
\begin{eqnarray}\label{FF}
y'f+x'\hat Hf =
-y'\frac{\delta\cal H}{\delta y}- \psi'\frac{\delta\cal H}{\delta\psi}=
\frac{1}{2i}(Fz' - \bar F\bar z').
\end{eqnarray}
Here $F = f +i\hat Hf$, function being analytic in the lower half-plane.
By dividing on $J$ and applying the projective operator one can find:
\begin{eqnarray}\label{FFF}
F = -\frac{2i}{z'}\hat P^-(
y'\frac{\delta\cal H}{\delta y} + \psi'\frac{\delta\cal H}{\delta\psi}).
\end{eqnarray}
Note that expression (\ref{FFF}) does not include time derivatives.
Now one can express $\psi_t$ from equations (\ref{2implicit}) and (\ref{3implicit}), using equations (\ref{1implicit}) and (\ref{FFF}).
After simple calculations we end up with the following equation:
\begin{eqnarray}\label{ReCo2}
\psi_t = -(\psi_u\hat H\frac{1}{J} +\frac{1}{J})\frac{\delta\cal H}{\delta\psi}
-\frac{1}{J}(x_u+\hat Hy_u)\frac{\delta\cal H}{\delta y}.
\end{eqnarray}
Equations (\ref{ReCo}) and (\ref{ReCo2}) can be written in the ``implectic''
form
\begin{eqnarray}\label{1of2}
\dot\psi &=& \hat R_{11}\frac{\delta\cal H}{\delta\psi}
-\hat R_{12}\frac{\delta\cal H}{\delta y}, \cr
\dot y &=& \hat R_{21}\frac{\delta\cal H}{\delta\psi}
\end{eqnarray}
Here
\begin{eqnarray}\label{Rij}
\hat R_{11} &=& -(\psi^\prime \hat H\frac{1}{J}
+\frac{1}{J}\hat H\psi^\prime),\cr
\hat R_{22} &=& 0,\cr
\hat R_{12} &=& \frac{1}{J}(x^\prime+\hat Hy^\prime), \cr
\hat R_{21}&=& (x^\prime -y^\prime\hat H)\frac{1}{J},\hat R_{21}=\hat R_{21}^+.
\end{eqnarray}
These formulae determine the Poisson structure on the functionals defined on
the real functions $y$ and $\psi$. Let $\alpha$ and $\beta$ is a pair
such functionals. Obviously
\begin{eqnarray}\nonumber
\{ \alpha, \beta \} = \int_{-\infty}^{\infty}
\{ \frac{\delta\alpha}{\delta\psi(x)}\hat R_{11}
\frac{\delta\beta}{\delta\psi(x)} -
\frac{\delta\alpha}{\delta\psi(x)}\hat R_{12}
\frac{\delta\beta}{\delta y(x)}+
\frac{\delta\alpha}{\delta y(x)}\hat R_{21}
\frac{\delta\beta}{\delta\psi(x)}\}dx.
\end{eqnarray}

\section{Basic equations in the complex form}

It is convenient to accomplish equation (\ref{1complex}) imposed to the
complex potential
\begin{eqnarray}\label{ComPot}
\Phi &=& \psi +i\hat H\psi = 2\hat P^{-}\psi,\\
\bar{\Phi} &=& \psi -i\hat H\psi = 2\hat P^{+}\psi\nonumber.
\end{eqnarray}
Note that
\begin{equation}\label{dHdpsi}
\frac{\delta \cal H}{\delta \psi} = 2\left (
\hat P^{+}\frac{\delta \cal H}{\delta \Phi}
+\hat P^{-}\frac{\delta \cal H}{\delta\bar\Phi}\right )
\end{equation}
\begin{equation}\label{dHdy}
\frac{\delta \cal H}{\delta y} = 2\left (
\hat P^{+}\frac{\delta \cal H}{\delta z}
+\hat P^{-}\frac{\delta \cal H}{\delta\bar z}\right ).
\end{equation}
Hence
\begin{eqnarray}\label{Uagain}
U = 4\hat P^-\{\frac{1}{J}
(\hat P^-\frac{\delta \cal H}{\delta\bar\Phi}
+\hat P^+\frac{\delta \cal H}{\delta\Phi})\}.
\end{eqnarray}
Now equation (\ref{1complex}) can be presented in the complex form.

Now one can apply operator $2\hat P^-$ to the equation (\ref{ReCo2})
and express $\frac{\delta \cal H}{\delta \psi}$ and
$\frac{\delta \cal H}{\delta y}$ by the use of (\ref{dHdpsi}) and
(\ref{dHdy}). We get closed equation for $\dot\Phi$. This equation can be
transformed to the following simple form ( see Appendix \ref{Appendix2} )
\begin{equation}\label{D2}
\dot\Phi = iU\Phi' - B -\cal P.
\end{equation}
Here
\begin{equation}\label{B}
B = -4i\hat P^-\{\frac{1}{J}
(\hat P^-(\bar\Phi^\prime\frac{\delta \cal H}{\delta\bar\Phi})
-\hat P^+(\Phi^\prime\frac{\delta \cal H}{\delta\Phi}))\}
\end{equation}
and
\begin{equation}\label{P}
{\cal P} = -4i\hat P^-\{\frac{1}{J}
(\hat P^-\bar z^\prime\frac{\delta \cal H}{\delta\bar z}
-\hat P^+ z^\prime\frac{\delta \cal H}{\delta z})\}.
\end{equation}

Equations (\ref{1complex}) and (\ref{D2}) compose close system of Hamiltonian equations written in the complex form. Poisson bracket in terms of $(z, \Phi)$
is discussed in Appendix \ref{ap:PoissonBraket}.

There is another form of complex equations.
Following to the article \cite{D2001} we introduce new variables:
\begin{eqnarray}\label{Dy2001}
R &=& \frac{1}{z'},\cr
V &=& i\frac{\partial\Phi}{\partial z} = i R\Phi'\nonumber.
\end{eqnarray}
In terms of $R$ and $V$ equations (\ref{1complex}) and (\ref{D2})
take a form:
\begin{eqnarray}\label{DYA}
\frac{\partial R}{\partial t} &=& i(UR' - RU'),\nonumber\cr
\frac{\partial V}{\partial t} &=& i(UV' -R(B+{\cal P})').
\end{eqnarray}

It is important to stress that equations (\ref{1complex}), (\ref{D2})
and (\ref{DYA}) are written for the functions which are analytical in
the lower half-plane (Im$w<0$).

We should stress once more that equations (\ref{DYA}) are just another form
of the general Hamiltonian equations (\ref{pair}). Meanwhile, this particular
form of the Hamiltonian equations with the very noncanonical Poisson brackets
is in our opinion the most convenient for both analytic and numeric study
of hydrodynamics with free surface. Finally we present implicit equations
(\ref{1implicit}), (\ref{2of2}) in the complex form. To do this we remember
that
\begin{eqnarray}\label{psiy}
x &=& {1\over 2} (z +\bar z), \hspace{2cm} y=\frac{1}{2i}(z-\bar z),\\
\psi &=& {1\over 2} (\Phi +\bar\Phi), \hspace{2cm}
\hat H\psi=\frac{1}{2i}(\Phi-\bar \Phi).
\end{eqnarray}
One can see that equations (\ref{1implicit}), (\ref{2of2}) can be presented
as follow
\begin{equation}\label{Comp1}
\hat P^-\left [z_t\bar z_u - \bar z_t z_u +
4i \frac{\delta \cal H}{\delta \bar\Phi}\right ] = 0.
\end{equation}
\begin{equation}\label{Comp2}
\Phi_t z_u - z_t\Phi_u + \hat P^-\left (\bar\Phi_t z_u - z_t \bar\Phi_u +
4i \frac{\delta \cal H}{\delta \bar z}\right ) = 0.
\end{equation}

\section{Hydrodynamics with a free surface}

All the machinery described above was designed to study different versions of
potential flows of ideal fluid with a free surface. We start syatematic
study of these models from the simplest one - "free" incompressible fluid
moving only under influence of force of inertia. In this case the Hamiltonian
is just kinetic energy of the fluid
\begin{eqnarray}\label{H1}
{\cal H} = T = -\frac{1}{2}\int_{-\infty}^{\infty}\psi\hat H\psi'du =
{i\over 8}\int_{-\infty}^{\infty}(\Phi +\bar\Phi)
(\Phi^\prime -\bar\Phi^\prime)du,
\end{eqnarray}
Now
\begin{eqnarray}\label{VarDer}
\frac{\delta \cal H}{\delta \psi} &=& -\hat H \psi,\\
\frac{\delta \cal H}{\delta \Phi} &=& \frac{i}{4}\Phi_u,\hspace{2cm}
\frac{\delta \cal H}{\delta \bar\Phi} = -\frac{i}{4}\Phi_u
\end{eqnarray}
Let us consider first the implicit equation. Equations (\ref{1implicit}) and
(\ref{2of2}) take the following form
\begin{equation}\label{quad1}
y_t x_u-x_t y_u = -\hat H \psi_u,
\end{equation}
\begin{equation}\label{quad2}
\psi_t x_u -x_t\psi_u -\hat H(y_u\psi_t-y_t\psi_u) =0.
\end{equation}
Equation (\ref{quad1}) can be rewritten as follow
\begin{equation}\label{quad1_complex}
z_t\bar z_u-\bar z_t z_u = -2i\hat H \psi_u.
\end{equation}
This is the step to complex form of implicit equations.
Full complex implicit equations are:
\begin{equation}\label{quad1_complex2}
z_t\bar z_u-\bar z_t z_u = \bar\Phi_u - \Phi_u,
\end{equation}
or after applying projector operator $\hat P^-$
\begin{equation}\label{quad1_complex3}
\hat P^- \left (z_t\bar z_u-\bar z_t z_u \right ) = - \Phi_u.
\end{equation}
Second equation is
\begin{equation}\label{quad2_omplex}
\Phi_t z_u - z_t\Phi_u +
\hat P^-\left (\bar\Phi_t z_u - z_t \bar\Phi_u \right ) = 0.
\end{equation}
The "transport velocity" $U$ is defined only by kinetic energy. For all
hydrodynamic model $U$ and $B$ are universal
\begin{eqnarray}\label{UandB}
U &=& -2 \hat P^-\left (\frac{1}{J}\hat H\psi_u \right ) =
i\hat P^-\left (\frac{\Phi_u-\bar\Phi_u}{|z-u|^2}\right ),\\
B &=& \hat P^-\left (\frac{|\Phi_u|^2}{|z_u|^2}\right ).
\end{eqnarray}
Thus "scalar" explicit equations read:
\begin{eqnarray}\label{Scalarexplicit}
\dot y &=& -(x' - y' \hat H)\frac{1}{J}\hat H\psi_u,\cr
\dot\psi &=&
\psi_u\frac{\hat H\psi_u}{J} +
\frac{1}{J}\hat H (\psi_u \hat H \psi_u).
\end{eqnarray}
The "complex" explicit equations has the standart form
\begin{eqnarray}\label{DYA3}
\dot z &=& iUz^\prime,\cr
\dot\Phi &=& iU\Phi' - B.
\end{eqnarray}
$U$ and $B$ are given by equations (\ref{UandB}).
Finally the Dyachenko equations for this simplest case read
\begin{eqnarray}\label{DYA2}
R_t &=& i(UR' - RU'),\cr
V_t &=& i(UV' -R B').
\end{eqnarray}
In $R$ and $V$ variables
\begin{eqnarray}\label{UHyd}
U = \hat P^-(R\bar V + \bar R V),
\end{eqnarray}
\begin{eqnarray}\label{BHyd}
B =  \hat P^-(V\bar V).
\end{eqnarray}
One more usefull form of "free" hydrodynamic equations stems from (\ref{FFF}).
Apparently
\begin{eqnarray}\label{FFF_1}
F = -\frac{i}{z'}\hat P^-(
\psi'\frac{\delta\cal H}{\delta\psi}) = \frac{i}{z'}\hat P^-(
\psi'\hat  H\psi')=\\
\frac{i}{2z'}\hat P^-(\mbox{Im}{\Phi'}^2) = \frac{{\Phi'}^2}{2 z'}.
\end{eqnarray}

From (\ref{2implicit}) and (\ref{3implicit}) we obtain important equation
\begin{equation}\label{BERNOULLI_C}
\dot\psi\bar z_u -\dot z \psi_u + \frac{\Phi_u^2}{2 z_u^2} = 0.
\end{equation}
This is the Bernoulli equation written in conformal variables.

\section{Dirichlet solutions}

"Free" hydrodynamic equations admit a remarkable class of solutions, found
first by L. Dirichlet in 19 century. In These solutions potential is a
quadratic function of coordinates, and the free surface is a curve of second
order. It is interesing to find these solutions in conformal variables. Let
us assume that
\begin{equation}\label{PhiQuadDirichlet}
\Phi = a(t) + \frac{b(t)}{2}z^2,
\end{equation}
$a(t)$ and $b(t)$ are real.
Then
$$\Phi' = bzz',$$
$$\psi = \frac{1}{2}(\Phi +\bar \Phi) = a + \frac{b}{4}(z^2+{\bar z}^2).$$
Equation (\ref{quad1_complex2}) reads now
\begin{equation}\label{1_quad}
\dot z \bar z' - \dot{\bar z} z' = b(\bar z \bar z' - z z').
\end{equation}
Equation (\ref{BERNOULLI_C}) takes the form
\begin{equation}\label{2_quad}
\left [\dot a +\frac{\dot b}{4}(z^2 +{\bar z}^2 \right ]\bar z' + \frac{bz}{2}(\dot z \bar z' - \dot{\bar z}z')
+\frac{b^2}{2}z^2z' = 0.
\end{equation}
Using equation (\ref{1_quad}) we realize that the terms proportional to $z'$
are cancelled. Then we cancel $\bar z$ and end up with the equation for the
quadric
\begin{equation}\label{quadric}
\dot a + \frac{\dot b}{4}(z^2 + {\bar z}^2) +\frac{b^2}{2}|z|^2 = 0.
\end{equation}
This equation describe the shape of the surfce. To find conditions for
$\dot a$, $\dot b$ we should solve the equation (\ref{1_quad}).

Suppose that $\dot b - b^2 > 0$. Now quadric (\ref{quadric}) is hyperbola
\begin{eqnarray}\label{hyperbola}
\frac{y^2}{A^2} &=& 1 + \frac{x^2}{B^2},\cr
A^2 &=& \frac{2\dot a}{\dot b - b^2}, \hspace{0.5cm} B^2 = \frac{2\dot a}{\dot b + b^2}
\end{eqnarray}
Equation (\ref{1_quad}) can be rewritten as follow
\begin{equation}\label{1_hyper}
\dot y x' - \dot x y' = -b(x y' + x'y).
\end{equation}
By plugging (\ref{hyperbola}) into (\ref{1_hyper}) one get
\begin{eqnarray}\label{2_hyper}
\dot A = -bA, \hspace{1cm} \dot B = bB,\cr
AB = \lambda^2 = const, B^2 = \frac{\lambda^4}{A^2}
\end{eqnarray}
Then
$$
A^2(\dot b - b^2) = B^2(\dot b + b^2)
$$
or
\begin{equation}\label{3_hyper}
\dot b( 1-\frac{\lambda^4}{A^4}) = b^2( 1+\frac{\lambda^4}{A^4})
\end{equation}
From (\ref{3_hyper}) one can see that the case $A^2 = B^2 = \lambda^2$ is marginal. If $A^2 < \lambda^2$, $\dot b <0$. in the opposite case $A^2 > \lambda^2$, $\dot b > 0$. By assuming that
$$
\dot b = \dot A\frac{db}{dA} = -Ab\frac{db}{dA},
$$
one get
\begin{equation}\label{4_hyper}
\frac{db}{dA} = -\frac{1}{A}\frac{A^4+\lambda^4}{A^4-\lambda^4}.
\end{equation}
Thus the equation can be easily integrated
\begin{equation}\label{5_hyper}
b = \pm\frac{FA}{|A^4-\lambda^4|^{1/2}} = -\frac{\dot A}{A},
\end{equation}
$F>0$ - is constant of integration.
Equation (\ref{5_hyper} can be integrated elliptical functions. In two limiting cases $A<<\lambda$ and  $A>\lambda$ it essentially simplifies.

Let $A|_{t=0} = A_0$.\\
If $A_0<<\lambda$\\
\begin{eqnarray}\label{6_hyper}
\dot A = -FA^2, \hspace{1cm} A = \frac{F^{-1}}{t+q}, \hspace{1cm} q= \frac{F}{A_0}.
\end{eqnarray}
If $A_0>>\lambda$\\
\begin{eqnarray}\label{7_hyper}
\dot A = F, \hspace{1cm} A = Ft + A_0.
\end{eqnarray}
Hence the "flat" hyperbola $A_0<\lambda$ becomes more flat, while the "sharp" hyperbola
$A_0>\lambda$ becomes more sharp as $t\rightarrow\infty$.

Let us introduce new variables $c$ and $\theta$ such that
$$
A = c\cos{\theta}, \hspace{1cm} B = c\sin{\theta}
$$
\begin{eqnarray}\label{8_hyper}
z &=& -\frac{ic}{2}(Re^{i\theta} +\frac{1}{R}e^{-i\theta}),
\end{eqnarray}
\begin{eqnarray}\label{9_hyper}
R &=& e^{\frac{2\theta}{\pi}arcsinh(w)} = (w+\sqrt{w^2+1})^{\frac{2\theta}{\pi}}.
\end{eqnarray}
Then
\begin{eqnarray}\label{10_hyper}
\dot R &=& \frac{2\dot\theta}{\pi}arcsinh(w)R = \frac{2\dot\theta}{\pi}(w+\sqrt{w^2+1})R,
\end{eqnarray}
\begin{eqnarray}\label{11_hyper}
R' &=& \frac{2\theta}{\pi}\frac{1}{\sqrt{w^2+1}}R,
\end{eqnarray}
\begin{eqnarray}\label{12_hyper}
z' &=& -\frac{ic}{\pi}\frac{1}{\sqrt{w^2+1}}(Re^{i\theta} +\frac{1}{R}e^{-i\theta}),
\end{eqnarray}
\begin{eqnarray}\label{13_hyper}
\dot z &=& -\frac{i}{2}\left [ (\dot c + ic\dot\theta)Re^{i\theta}
 + (\dot c - ic\dot\theta)\frac{1}{R}e^{-i\theta} \right ]
-\frac{ic\dot\theta}{\pi}arcsinh(w) (Re^{i\theta} +\frac{1}{R}e^{-i\theta}),
\end{eqnarray}
Now remember that
\begin{eqnarray}\label{14_hyper}
AB &=& c^2\sin{\theta}\cos{\theta} = \frac{1}{2}c^2\sin{2\theta}.
\end{eqnarray}
By differentiating (\ref{14_hyper}) one get
\begin{eqnarray}\label{15_hyper}
\dot c &=& -\frac{c\cos{2\theta} }{\sin{2\theta} }\dot\theta.
\end{eqnarray}
Finally we got
\begin{eqnarray}\label{16_hyper}
\dot z &=& \frac{ic\dot\theta}{2} \frac{1}{\sin{2\theta}}
(Re^{i\theta} +\frac{1}{R}e^{-i\theta})
-\frac{2}{\pi}arcsinh(w)(Re^{i\theta} +\frac{1}{R}e^{-i\theta}),
\end{eqnarray}
or
\begin{eqnarray}\label{17_hyper}
\dot z &=& \frac{dot\theta}{\sin{2\theta}}\bar z
+\frac{\pi\dot\theta}{\theta}\sqrt{w^2+1}arcsinh(w)z'.
\end{eqnarray}
As we know
\begin{eqnarray}\label{18_hyper}
b &=& \frac{\dot A}{A} = \frac{-\dot c\cos{\theta} +c\dot\theta\sin{\theta}}{c\cos{\theta}}=
\frac{c\dot\theta}{\sin{2\theta}}.
\end{eqnarray}
Tj get the last result we need equation (\ref{4_hyper}). Now by plugging (\ref{13_hyper}), (\ref{17_hyper}), (\ref{18_hyper}) into (\ref{1_quad}) we see that this equation is automatically satisfied. Now we can find the transport velocity $U$. According to (\ref{DYA3})
\begin{equation}\label{19_hyper}
U = -\frac{i\dot z}{z'}.
\end{equation}
By the use of (\ref{12_hyper}) and (\ref{17_hyper}) we get
\begin{equation}\label{20_hyper}
U = -\frac{i\pi\dot\theta}{\theta} \sqrt{w^2+1} \left [
\frac{1}{2\sin{2\theta}} \frac{Re^{i\theta} +\frac{1}{R}e^{-i\theta}} {Re^{i\theta} -\frac{1}{R}e^{-i\theta}} +\log{(w+\sqrt{w^2+1})}
\right ].
\end{equation}
Note that $z$ has only one branch point at $w=i$. It is analytic everywhere exept the cut along the imaginary axis $(i,i\infty)$. The same statement is correct for all other analytic in the lower half-plane functions
$$
R = \frac{1}{z'}, V= bz, B = \hat P^-(V\bar V).
$$
Apparently
\begin{equation}\label{21_hyper}
R = \frac{i\pi}{c\theta}\sqrt{w^2+1}\frac{1}{Re^{i\theta} -\frac{1}{R}e^{-i\theta}}.
\end{equation}
Note that
\begin{equation}\label{22_hyper}
R|_{w=i} = 0,\hspace{1cm} U|_{w=i} = 0.
\end{equation}
To find $B$ we again can use equation (\ref{DYA3}).

\section{This is the end...}

To apply the developed general theory to concrete physical system one should
specify the Hamiltonian function $\cal H$. We will study the following
class of Hamiltonian functions:
\begin{eqnarray}\label{H1234}
{\cal H} = {\cal H}_1 + {\cal H}_2 + {\cal H}_3 +{\cal H}_4.
\end{eqnarray}
Here

\begin{eqnarray}\label{H2}
{\cal H}_2 ={i\alpha\over 8}\int_{-\infty}^{\infty}(z' +\bar z'-2)
(z -\bar z)du = -\frac{\alpha}{2}\int_{-\infty}^{\infty}y(x'-1)du,
\end{eqnarray}
\begin{eqnarray}\label{H3}
{\cal H}_3 = -\frac{g}{16}\int_{-\infty}^{\infty}(z-\bar z)^2 (z'+\bar z')du
=\frac{g}{2}\int_{-\infty}^{\infty}y^2x'du,
\end{eqnarray}
\begin{eqnarray}\label{H4}
{\cal H}_4 = \sigma\int_{-\infty}^{\infty}\left(\sqrt{z'\bar z'}
-\frac{(z'+\bar z')}{2}\right)du = \sigma\int_{-\infty}^{\infty}
\left(\sqrt{{y'}^2 +{x'}^2} - x'\right)du.
\end{eqnarray}
Hamiltonian ${\cal H}_1$ is the kinetic energy of the ideal
incompressible fluid. It depends on $\Phi$ only, not on $z$.
The others, ${\cal H}_2$, ${\cal H}_3$ and ${\cal H}_4$, are
different forms of potential energy depending on $z$ only.
Hamiltonians ${\cal H}_3$ and ${\cal H}_4$ are
the components of the potential energy in the presence of gravity
acceleration $g$ and surface tension $\sigma$. Hamiltonian ${\cal H}_2$
has more sophisticated physical interpretation (see \cite{Zub2000}).
For $\alpha < 0$ it is
a potential energy of the dielectric fluid with ideally conducting
free surface, placed into the electric field. For $\alpha > 0$ this is
the potential energy of the same fluid containing inside the magnetic field.
More detailed information is in the Apendix B.

For the Hamiltonian function (\ref{H1234})
\begin{eqnarray}\label{UHyd}
U = i\hat P^-\{\frac{1}{\bar z^\prime}\frac{\Phi^\prime}{z^\prime}
- \frac{1}{z^\prime}\frac{\bar\Phi^\prime}{\bar z^\prime}\} =
\hat P^-(R\bar V + \bar R V),
\end{eqnarray}
\begin{eqnarray}\label{BHyd}
B = \hat P^-\{\frac{\Phi^\prime}{z^\prime}
\frac{\bar\Phi^\prime}{\bar z^\prime}\} = \hat P^-(V\bar V).
\end{eqnarray}
In the absence of a potential energy ${\cal P}=0$ (see (\ref{P})). In the
presence of a potential energy
\begin{eqnarray}\label{P234}
{\cal P} = {\cal P}_2 + {\cal P}_3 +{\cal P}_4.
\end{eqnarray}
Here
\begin{eqnarray}\label{P2}
{\cal P}_2 = \alpha\hat P^-(R\bar R-1),
\end{eqnarray}
\begin{eqnarray}\label{P3}
{\cal P}_3 = -ig(z - w),
\end{eqnarray}
\begin{eqnarray}\label{P4}
{\cal P}_4 = 2\sigma\hat P^-(Q'\bar Q - \bar{Q'}Q), \hspace{2cm} Q=\sqrt{R}.
\end{eqnarray}

In the absence of the surface tension motion equations (\ref{DYA}) take
the following remarkably simple form:
\begin{eqnarray}\label{DYA22}
R_t &=& i(UR' - RU'),\cr
V_t &=& i(UV' -R\hat P^-(V\bar V +\alpha R\bar R)') +g(R-1).
\end{eqnarray}
Equations (\ref{DYA22}) with (\ref{UHyd}) are cubic with respect to
the unknown functions $R$ and $V$. It makes them very suitable for
numerical simulation.

Let us put now $\alpha = 0$, and include into consideration the
surface tension. The first equation of (\ref{DYA22}) is not changed,
while the second takes the form:
\begin{eqnarray}\label{Vsig}
V_t = i(UV' - R \hat P^-(V\bar V)' +g(R-1) ) -2\sigma R
\hat P^-(Q'\bar Q - \bar Q' Q)',
\end{eqnarray}
For function $Q$ equations (\ref{DYA22}) became quartic nonlinear:
\begin{eqnarray}
\label{quartic}
Q_t &=& i(UQ' - {1\over 2}U'Q), \cr
V_t &=& i(UV' - Q^2 \hat P^-(V\bar V)') + g(Q^2-1)
-2\sigma Q^2\hat P^-(Q'\bar Q - \bar Q' Q)',
\end{eqnarray}
and $U$ is equal to:
$$ U = \hat P^-(V\bar Q^2 + \bar V Q^2).$$
Equations (\ref{DYA22}) keep the same form for other boundary conditions.
Just for different boundary conditions we need to replace Hilbert
transformation by
some other operator. For instance for periodic boundary conditions one has
to use Schwartz transformation instead of Hilbert. Another important case,
fluid of finite depth is discussed in~\cite{DKZ96}.

Note that equations include now only derivatives of conformal mapping
and complex velocity potential. As regards integrals of motion they acquire
more complicated form. But if one  restores complex velocity potential
$$\Phi = -i \int \frac{V}{R}dw$$
then the
kinetic energy is equal to
$$\Ha_1 = -\int_{-\infty}^{\infty} Re(\Phi) Im(\Phi')du$$
and momenta (given by (\ref{momenta})) are now equal to:
$$P_y = \int_{-\infty}^{\infty} Re(\Phi) Re(\frac{1}{R})du, \,\,\,\,
P_x = \int_{-\infty}^{\infty} Re(\Phi) Im(\frac{1}{R})du.$$

\section{New integrals of motion}

Equations (\ref{DYA22}) have some remarkable properties. Suppose that
$z(w)$ and $\Phi(w)$ have at $w = \lambda$ a logarithmic branch-point
($\mbox{Im}\lambda >0$):
\begin{eqnarray}\label{log}
z &=& a\log{(w-\lambda)} +w,\cr
\Phi &=& b\log{(w-\lambda)}.
\end{eqnarray}
At this point
\begin{eqnarray}\label{poles}
z' &=& \frac{a}{w-\lambda} +1,\cr
\Phi' &=& \frac{b}{w-\lambda}
\end{eqnarray}
and
$$V = i\frac{\Phi'}{z'} \simeq i\frac{b}{a},\hspace{2cm}
R\simeq \frac{1}{c}(w-\lambda).$$
One can see that both $R$ and $V$ are regular in the neighborhood
of point $\lambda$. the same statement is correct for $B$, ${\cal P}_2$
and ${\cal P}_3$. Note that it is not correct for ${\cal P}_4$. Indeed,
near $w = \lambda$
$$
{\cal P}_4 \simeq \frac{\sigma}{\sqrt{w-\lambda}}.
$$

Regularity of $U$ and $B$ results in the following important
consequence. Suppose that functions $R$ and $V$ are regular not only
in the lower half-plane, but also in some
domain above the real axis. And in this domain $R$ has simple zeroes
at the points
$$\lambda_1, \lambda_2,\dots \lambda_n.$$
In the vicinity of each zero
\begin{eqnarray}\nonumber
R &\simeq& a_n (w-\lambda_n),\cr
\mbox{and}\hspace{1cm} V &\simeq& V_n.
\end{eqnarray}

If $R$ and $V$ satisfy the equations (\ref{DYA22}), then
\begin{eqnarray}\label{int1}
\frac{da_n}{dt} = 0,\hspace{2cm} \frac{dV_n}{dt} = g
\end{eqnarray}
In the other words, coefficients $a_n$, which are nothing but
logarithmic residues of $z$, are constant of motion of the equations
(\ref{DYA22}). In the absence of gravity logarithmic residues of $\Phi$
are constant of motion either. In presence of gravity they are linear
functions of time
$$ V_n = {V_n}_0 + gt$$
and
$$b_n = -i a_n V_n = - i a_n ({V_n}_0 + gt).$$

It is important to find the Poisson's bracket between new integrals
of motion. As far as $a_n$ are completely defined by $z$, so that
$$\frac{\delta a_n}{\delta\Phi}=\frac{\delta a_n}{\delta\bar\Phi}=0,$$
one can conclude:
\begin{eqnarray}\label{anam}
\{a_n,a_m\}.
\end{eqnarray}
Poisson's brackets
$$
\{a_n,b_m\},\hspace{1.5cm} \mbox{and}\hspace{1.5cm} \{b_n,b_m\}
$$
are still unknown.

So far we have discussed only simple zeroes of $R$. In fact, one can
release this condition and consider zeroes of higher order. One can
assume that in the vicinity of $\lambda$
\begin{eqnarray}\label{mzeroes}
U &\simeq& u_0 + (w-\lambda)u_1,\cr
R &\simeq& (w-\lambda)^n \tilde R.
\end{eqnarray}
Plugging (\ref{mzeroes}) into the (\ref{DYA22}) one can find
\begin{eqnarray}\label{tildeR}
\dot{\tilde R} &=& i(U\tilde R'  -\tilde RU' + n u_1 \tilde R),\cr
\dot\lambda &=& -u_0.
\end{eqnarray}
This assumption (\ref{mzeroes}) is structurally stable. One can assume
again that $V$, $U$, $B$ and $\cal P$ are regular at $w=\lambda$.
For
$$
V_n = V\vert_{w=\lambda_n}
$$
one can obtain the following formula:
\begin{eqnarray}\label{dotVg}
\frac{dV_n}{dt} = g.
\end{eqnarray}
In this case both $z'$ and $\Phi'$ have a pole of $n$-th order at
the point $w=\lambda$:
\begin{eqnarray}\label{mpoles}
z'&=&\frac{a_{-n}}{(w-\lambda)^n}+\dots+\frac{a_{-1}}{(w-\lambda)}+\dots,\cr
\Phi'&=&\frac{b_{-n}}{(w-\lambda)^n}+\dots+\frac{b_{-1}}{(w-\lambda)}+\dots.
\end{eqnarray}
Plugging (\ref{mpoles}) into the equations (\ref{DYA3}), one can obtain:
\begin{eqnarray}\label{mpol}
\frac{d}{dt}a_{-1} = 0, \hspace{2cm} \frac{d}{dt}b_{-1} = a_{-1}g
\end{eqnarray}
Moreover, (\ref{dotVg}) reads:
\begin{eqnarray}\label{mVg}
\frac{d}{dt}\frac{b_{-n}}{a_{-n}} = g.
\end{eqnarray}

Formulae (\ref{mpol}) can be written as follow:
\begin{eqnarray}\label{INT12}
\frac{d}{dt}\oint_\Gamma z^\prime dw =0,\hspace{1.5cm}
\frac{d}{dt}\oint_\Gamma  \Phi^\prime dw=g\oint_\Gamma z^\prime dw.
\end{eqnarray}
In (\ref{INT12}) one integrates along small contour around $\lambda$.	

Formulae (\ref{dotVg}) and (\ref{INT12}) do not include the parameter
$\alpha$. One can think that they could be extended to a more broad
class of the Hamiltonian functions. This is actually not true. Only a very
special Hamilton functions keep $U$, $B$ and $\cal P$ analytic at the points
where $z'$ and $\Phi'$ have poles. In a general case singularities in
$z'$ and $\Phi'$ generate singularities in $U$, $B$ and $\cal P$, and
formulae (\ref{dotVg}), (\ref{mpol}), (\ref{mVg}) and (\ref{INT12})
are violated.

This phenomenon can be traced for the case when we include into
consideration surface tension. If $R$ has a simple zero
\begin{eqnarray}\label{zerooo}
R\simeq (w-\lambda),
\end{eqnarray}
the expression for surface tension term in (\ref{quartic}) is
\begin{eqnarray}\label{br}
-2\sigma R\hat P^-(Q'\bar Q - \bar Q' Q)'\simeq
\frac{c}{(w-\lambda)^{1\over 2}}.
\end{eqnarray}
It means that analyticity of $V$ is immediately violated. The situation
can be fixed if $Q = \sqrt{R}$ is analytic functions, having zero of
any arbitrary integer order. one can see that the assumption
\begin{eqnarray}\label{Qz}
Q = (w-\lambda)^n\tilde Q
\end{eqnarray}
is compatible with equations (\ref{quartic}).

Expression (\ref{Qz}) for $Q$ implies that
\begin{eqnarray}\label{Rz}
R = (w-\lambda)^{2n}\tilde R.
\end{eqnarray}
Thus $R$ has zero of an even order.

In conclusion of this chapter we should stress that even in the absence
of gravity integrals (\ref{dotVg}) and (\ref{INT12}) cannot be
interpreted as Casimirs for some degenerated Poisson's bracket. Just
due to the fact of existence of the variational principle and the
symplectic structure, the Poisson's bracket in this case is not
degenerated, and no Casimirs do exist.

\section{Connection to the LGE equation}
In the articles \cite{DZ96}, \cite{ZD96PD} we have found that in certain
situations the dynamics of a free-surface fluid can be described by
the Laplace growth Equation (LGE), which is known in hydrodynamics
since 1945 (\cite{PK45}, \cite{G45}). We show now that the LGE appears
in a natural way in the framework of our formalism.

Let us suppose that the fluid flow has two scales - global large
scale with the complex potential $\Psi_0$, and small scale with the potential
$\delta\psi$. Another words:
\begin{equation}\label{separation}
\Psi = \Psi_0 +\delta\psi.
\end{equation}
Let us denote
\begin{equation}\label{V0}
V_0 = \hat H\Psi'_0.
\end{equation}
Substituting (\ref{separation}) into kinetic energy (\ref{H1}), we neglect
the quadratic term
\begin{eqnarray}\nonumber
-\frac{1}{2}\int_{-\infty}^{\infty}\delta\psi\hat H\delta\psi'du.
\end{eqnarray}
Then
\begin{eqnarray}\label{LGE1}
{\cal H}_1 &=& \frac{1}{2}\int_{-\infty}^{\infty}\Psi_0 V_0du +
\int_{-\infty}^{\infty}V_0\delta\psi du, \cr
\frac{\delta{\cal H}_1}{\delta\Psi} &=& V_0(u,t).
\end{eqnarray}
Here $V_0(u,t)$ is a diven ``slow'' function of $u$ and $t$.

Now
\begin{eqnarray}\label{ULGE}
U = \hat P^-\{\frac{V_0}{|z'|^2}\}
\end{eqnarray}
and equation (\ref{reverse1}) takes a closed form:
\begin{eqnarray}\label{LGE2}
\mbox{Im}\{\dot z\bar z'\} = V_0.
\end{eqnarray}

In the absence of a potential energy $V_0$ can be put constant. In this
case equation (\ref{LGE2}) is exactly Laplace Growth Equation. In presence
of potential energy $V_0$ should be at least a function of time. In presence
of gravity
\begin{eqnarray}\label{LGEGRAV}
\dot V_0 &=& -g, \cr
V_0 &=& -gt + \mbox{constant}.
\end{eqnarray}

This approximation to the Laplace Growth Equation derived-ed from
hydrodynamics equation was obtained in \cite{DZ96}, \cite{ZD96PD}.

\section{Cuts - possible type of solution}

Suppose both $R$ and $V$ have cuts along imaginary axis in the upper
half-plane
\begin{equation}\label{GcutRV}
R(w,t) = 1 -\int_{a(t)}^{b(t)}\frac{{\it r}(s,t)ds}{s+iw},\hspace{1cm}
V(w,t) = \int_{a(t)}^{b(t)}\frac{{\it v}(s,t)ds}{s+iw}
\end{equation}
Here ${\it r}(s,t)$ and ${\it v}(s,t)$ are real functions of real argument
$s$ given in the interval $[a(t),b(t)]$, and
\begin{equation}\label{zeroends}
{\it r}(a,t) = {\it r}(b,t) = {\it v}(a,t) = {\it v}(b,t) =0.
\end{equation}
We show now that the anzats (\ref{GcutRV}) is the solution of the equations
(\ref{DYA2}).
Let us calculate complex transport velocity $U$.
$$U = \hat P^-\left\{R\bar V + \bar R V\right\}.$$
Anzats (\ref{GcutRV}) results in the following expression for $U$,
\begin{equation}\label{cutU}
U = \int_a^b\frac{{\it v}(s,t)}{s+iw}ds - \int_a^b\int_a^b
\frac{{\it r}(s',t){\it v}(s,t)+{\it r}(s,t){\it v}(s',t)}{s'+s}
\frac{ds'ds}{s+iw}
\end{equation}
It is convenient to introduce the notation:
\begin{eqnarray}\label{pro}
F(s,t) &=& \int_a^b\frac{{\it r}(s',t)}{s'+s}ds',\cr
G(s,t) &=& \int_a^b\frac{{\it v}(s',t)}{s'+s}ds'.
\end{eqnarray}
Then $U$ is very similar to $V$, namely
\begin{equation}
U = \int_a^b\frac{{\it u}(s,t)ds}{s+iw}.
\end{equation}\nonumber
Here
\begin{equation}\label{italicu}
{\it u}(s,t) = {\it v}(s,t) -
F(s,t){\it v}(s,t)-{\it r}(s,t)G(s,t)
\end{equation}
It is obvious that ${\it u}(s,t)$ is also equal to zero at the ends of
the interval $[a(t),b(t)]$. Similar formula one get for $B$:
\begin{equation}\label{italicB}
B=\hat P^-\left\{V\bar V\right\} = \int_a^b\frac{{\it b}(s,t)}{s+iw}ds
\end{equation}
Here
\begin{equation}\label{italicb}
{\it b}(s,t) =  {\it v}(s,t)G(s,t)
\end{equation}
And again, ${\it b}(s,t)$ is equal to zero at the ends of the
interval $[a(t),b(t)]$.

Now, let us calculate r.h.s of the equation for $R$,
\begin{eqnarray}\nonumber
&i&\left(UR' - U'R\right) =
\int_a^b\int_a^b\frac{{\it u}'(s,t){\it r}(s',t) - {\it r}'(s,t){\it u}(s',t)}
{(s+iw)(s'+iw)}ds'ds - \int_a^b\frac{{\it u}'(s,t)}{s+iw}=\cr
&=&\int_a^b\int_a^b\frac{{\it u}'(s,t){\it r}(s',t) +{\it u}'(s',t){\it r}(s,t)
- {\it r}'(s,t){\it u}(s',t) - {\it r}'(s',t){\it u}(s,t)}{s'-s}
\frac{ds'ds}{s+iw} - \int_a^b\frac{{\it u}'(s,t)}{s+iw}ds.
\end{eqnarray}
If we introduce functions
\begin{eqnarray}\label{ACPQ}
A(s,t) &=& P.V.\int_a^b\frac{{\it r}(s')}{s'-s}ds,\cr
C(s,t) &=& P.V.\int_a^b\frac{{\it v}(s')}{s'-s}ds,\cr
P(s,t) &=& -P.V.\int_a^b\frac{{\it r}(s')G(s',t)+
{\it v}(s')F(s',t)}{s'-s}ds,\cr
Q(s,t) &=& P.V.\int_a^b\frac{{\it v}(s')G(s',t)}{s'-s}ds
\end{eqnarray}

then equation for $R$ reads:
\begin{equation}
\dot{\it r}(s,t) =-{\it u}'(s,t)\hat{\it r}(s,t)-\hat{\it u}'(s,t){\it r}(s,t)
+ {\it r}'(s,t)\hat{\it u}(s,t) + \hat{\it r}'(s,t){\it u}(s,t) +{\it u}'(s,t)
\end{equation}
Doing the same thing for the equation for $V$ one can obtain:
\begin{equation}
\dot{\it v}(s,t) ={\it u}(s,t)\hat{\it v}'(s,t)+\hat{\it u}(s,t){\it v}'(s,t)
+ {\it b}'(s,t)\hat{\it r}(s,t) + \hat{\it b}'(s,t){\it r}(s,t) -{\it b}'(s,t)
\end{equation}
or with new notation:
\begin{eqnarray}\label{Znotation}
\dot{\it r}(s,t)&+&((1-A)G-C-P){\it r}'(s,t) - (1-A)(1-F){\it v}'(s,t) =\cr
&=& ((A-1)G'-A'G-C'-P'){\it r}(s,t)+((1-F)A'-(1-A)F'){\it v}(s,t),\cr
\dot{\it v}(s,t)&+&((1-A)G-C-P){\it v}'(s,t)=\cr
&=& (Q'-GC'){\it r}(s,t)+((1-F)C'-(1-A)G'){\it v}(s,t)
\end{eqnarray}

\section{Equations for spectral density on the cuts for implicit equations}
 Here we will consider the hydrodynamics equations for $y$ and $\psi$ given
in the
implicit form (\ref{2of2}). Actually we will use instead of $y$ and $\psi$
complex functions $z$ and $\Phi$, and the equations (\ref{2of2}) can be
rewritten for them:
\begin{equation}\label{dotz}
\dot z\bar z' - \dot{\bar z} z' = \bar\Phi' - \Phi'
\end{equation}
\begin{equation}\label{dotPhi}
\dot\Phi z'-\Phi'\dot z +\hat P^-(\dot{\bar\Phi} z' -\bar\Phi' \dot z) +
\dot{\bar\Phi}\bar z'-\bar\Phi'\dot{\bar z} +
\hat P^+(\dot\Phi\bar z' -\Phi' \dot{\bar z}) =0.
\end{equation}
Let us make the following hypothesis. Suppose
that both $z$ and $\Phi$ are analytic functions in the lower half-plane
and both have the only singularity in the upper half-plane, namely
a cut on the imaginary axis:
$$\lambda(t) < v < \infty,$$
$\lambda(t)$ is some unknown function of time. Let us denote the coordinate
along the cut by $s$, so that
$$w = u+iv = u +is.$$
Now we introduce the spectral density on the cut for conformal mapping
$z$ and potential $\Phi$:
\begin{eqnarray}\label{DensZ}
z &=& \frac{1}{2\pi i}\int_{\lambda}^{\infty}
\frac{\rho (s')}{s'+iw}ds',\cr
\Phi &=& \frac{1}{2\pi i}\int_{\lambda}^{\infty}
\frac{i\phi (s')}{s'+iw}ds'.
\end{eqnarray}
Let us calculate $z$ on the cut from  right and from the left or,
another words, for
$$w = is \pm \epsilon,\hspace{1cm} \epsilon\rightarrow 0.$$
Then
\begin{equation}\label{ZPM}
z^{\pm} = \frac{1}{2\pi i}P.V.\int_{\lambda}^{\infty}
\frac{\rho (s')}{s'-s\pm i\epsilon}ds' \mp{1\over 2}\rho (s).
\end{equation}
Then one can see that
$$\rho(s) = z^- - z^+.$$
Let us introduce two more functions:
\begin{equation}\label{fofs}
f(s) = \frac{1}{2\pi}P.V.\int_{\lambda}^{\infty}
\frac{\rho (s')}{s'-s}ds' =-{i\over 2}(z^+ + z^-),
\end{equation}
and
\begin{equation}\label{Aofs}
A(s) = \frac{1}{2\pi}\int_{\lambda}^{\infty}
\frac{\rho (s')}{s'+s}ds'.
\end{equation}
Then
\begin{equation}\label{barzofs}
\bar z(s) = -\frac{1}{2\pi i}\int_{\lambda}^{\infty}
\frac{\rho (s')}{s'+s} = i A(s).
\end{equation}

Here is some useful notations:
\begin{eqnarray}\label{NOTA}
\frac{\partial}{\partial w} &=& \frac{\partial}{i\partial s} =
-i \frac{\partial}{\partial s},\cr
\bar z_u &=& \frac{\partial}{\partial w}\bar z = A_s,\cr
z_u &=& {1\over 2} \frac{\partial}{\partial w}(z^- +z^+) = f_s,\cr
\bar z_t &=& i A_t, \hspace{1cm}\bar z_t z_u \rightarrow A_t A_s.
\end{eqnarray}

Now let us consider spectral density for potential $\Phi$:
For values of $\Phi$ on the right and left sides of the cut one can
easily derive the following relations:
\begin{eqnarray}\label{densp}
\Phi^- - \Phi^+ &=& i\phi(s),\cr
{1\over 2}(\Phi^- + \Phi^+) &=&
\frac{1}{2\pi}P.V.\int_{\lambda}^{\infty}
\frac{\phi (s')}{s'-s}ds'.
\end{eqnarray}
And similar to what we just did for $z$, let us use the notation:
\begin{eqnarray}\label{gofs}
g(s) &=& \frac{1}{2\pi}P.V.\int_{\lambda}^{\infty}
\frac{\phi (s')}{s'-s}ds',\cr
B(s) &=& \frac{1}{2\pi}\int_{\lambda}^{\infty}
\frac{\phi (s')}{s'+s}ds'.
\end{eqnarray}
Then the following relations hold:
\begin{eqnarray}\label{barofPhi}
\bar \Phi(s) &=& \frac{1}{2\pi}\int_{\lambda}^{\infty}
\frac{\phi (s')}{s'+s}ds' = B(s),\cr
\frac{\partial}{\partial w}(\Phi^- -\Phi^+) &=&
\frac{\partial\phi}{\partial s},\cr
{1\over 2}(\Phi^- +\Phi^+) &=& g(s).
\end{eqnarray}

Now we can easily rewrite the equation for $\dot z$ (\ref{dotz}):
\begin{eqnarray}\label{one1}
\rho_t (1+A_s) - \rho_s A_t = -\phi_s.
\end{eqnarray}
To write down the equation for $\dot\Phi$ (instead of (\ref{dotPhi}))
one should use
the following useful formula:
$$A^-B^- - A^+B^+ = {1\over 2}[
(A^- -A^+)(B^- +B^+)+(A^- +A^+)(B^- -B^+)].$$
The the equation (\ref{dotPhi}) takes the form:
\begin{eqnarray}\label{two2}
\phi_t(1-f_s) + f_t\phi_s + \rho_t g_s -g_t\rho_s
+\rho_t B_s -\rho_s B_t =0,
\end{eqnarray}
or if we introduce new function
$$C(s) = g(s) +B(s) = {1\over\pi}P.V.\int_{\lambda}^{\infty}
\frac{\phi (s')s'}{s'^2-s^2}ds'$$
it is equal to:
\begin{eqnarray}\label{two22}
\phi_t(1-f_s) + f_t\phi_s + \rho_t C_s  -\rho_s C_t=0.
\end{eqnarray}

These equations (\ref{one1}) and (\ref{two22}) are the basic equations
describing the evolution of the densities on the cut.

Let us write them in the divergent form:
\begin{eqnarray}\label{divergent}
\frac{\partial}{\partial t}\left[\rho(1+A_s)\right] &=&
\frac{\partial}{\partial s}(-\phi +\rho A_t),\cr
\frac{\partial}{\partial t}\left[\phi(1-f_s)+\rho C_s\right] &=&
\frac{\partial}{\partial s}(-\phi f_t + \rho C_t)
\end{eqnarray}
Assuming that spectral density of the potential and conformal mapping
are equal to zero at the ends of the cut, we obtain the integral
of motion (for mass or the fluid and its vertical momentum):
\begin{eqnarray}
\int_{\lambda}^{\infty}\rho(1+A_s)ds = M,\cr
\int_{\lambda}^{\infty}\left[\phi(1-f_s)+\rho C_s\right]ds = P.
\end{eqnarray}
For the kinetic energy ${\cal H}$
\begin{equation}\label{kinen}
{\cal H} ={i\over 8}\int_{-\infty}^{\infty}(\Phi +\bar\Phi)
(\Phi^\prime -\bar\Phi^\prime)du,
\end{equation}
one can easily obtain the following formula:
\begin{eqnarray}
{\cal H} = -\frac{1}{4}\int_{\lambda}^{\infty}\phi A_s ds.
\end{eqnarray}

\section{Approximation for narrow cuts}
Suppose that cuts for $R$ and $V$ are far from the real axis, namely
their width $(b-a)$ is much less then the distance to the real axis
$$(b-a)<< a.$$
Then one can approximate
$$U =\hat P^-(\bar V R + V\bar R)$$
as
\begin{equation}\label{apprU}
U \simeq V_c R + V R_c - V_c,
\end{equation}
here $V_c$ is the value of $\bar V$ at the some point on the narrow
cut of $R$, and $R_c$ is the value of $\bar R$ on the narrow cut of $V$.
The last term in (\ref{apprU}) appears due to asymptotic of $R$ at infinity.

Here is the ground for
this approximation. Both $\bar R$ and
$\bar V$ have singularities in the lower half-plane, at the complex
conjugate points with respect to $R$ and $V$. If we consider
narrow cut (at the same place for $R$ and $V$), that means we assume
$V\simeq V_c$ and $R\simeq R_c$
being time dependent only.
This assumption allows us to get the approximate expression for
$B$ also:
\begin{equation}\label{apprB}
B = \hat P^-(\bar V V)\simeq V_c V.
\end{equation}
It should be mentioned here that for the limiting case of infinitely
narrow cur (it is nothing but pole) the approximation is exact.

Substituting (\ref{apprU}) and (\ref{apprB}) into the (\ref{DYA2})
we end up with the following equations:
\begin{equation}\label{2quadR}
\dot R+iV_cR' = iR_c(VR'-V'R)
\end{equation}
\begin{equation}\label{2quadV}
\dot V+iV_cV' = iR_c(VV').
\end{equation}
In the moving framework
$$\chi = w -i\int_0^t V_cdt$$
the equations (\ref{2quadR}) and (\ref{2quadV}) read:
\begin{equation}\label{quadR}
\dot R = iR_c(VR'-V'R)
\end{equation}
\begin{equation}\label{quadV}
\dot V = iR_c(VV'),
\end{equation}
where space derivative is now with respect to $\chi$.

It is remarkable that we derive complex Hopf equation (\ref{quadV}). If
we introduce here new time $\tau(t)$, so that
\begin{equation}\label{newtime}
\dot \tau(t) = R_c(t)
\end{equation}
and
\begin{equation}\label{neww}
\chi = w -i\int_0^\tau\frac{V_c}{R_c}d\tau.
\end{equation}
Recall that $R = \frac{1}{z'}$, than we finally get the following set
of quadratic equations:
\begin{equation}\label{quadZ}
z_{\tau} = iVz'
\end{equation}
\begin{equation}\label{quadV2}
V_{\tau} = iVV'.
\end{equation}
These equations are Hamiltonian ones, with the Hamiltonian
\begin{equation}\label{hamil0}
{\cal H} = \frac{i}{2}\int_{-\infty}^{\infty}V^2 z'd\chi
\end{equation}
and canonical variables $z$ and $V$,
\begin{equation}\label{varia}
z_{\tau} = \frac{\delta{\cal H}}{\delta V},\hspace{2cm}
V_{\tau} = -\frac{\delta{\cal H}}{\delta z}
\end{equation}
Note that $V$ and $z'$ are analytic function in the lower half-plane, and
$\cal H$ is equal to zero.

Equations (\ref{quadZ}) and (\ref{quadV2}) can be solved by the method
of characteristics.

Let us consider the following initial value problem for equations
(\ref{quadZ}) and (\ref{quadV2}):
\begin{eqnarray}\label{IVP}
V(\chi,\tau)\vert_{\tau=0} &=& \frac{A}{\lambda +i\chi},\cr
z(\chi,\tau)\vert_{\tau=0} &=& \chi.
\end{eqnarray}
Here $A$ and $\lambda$ are real positive constant. General solution
of the Hopf's equation (\ref{quadV2}) is given by:
$$F(V(\chi,\tau)) = i\chi - \tau V(\chi,\tau).$$
Initial condition (\ref{IVP}) defines the function $F(V))$:
$$F(V) = \frac{A}{V} - \lambda,$$
and we end up with the quadratic equation for $V(\chi,\tau)$:
$$\tau V^2 -(\lambda +i\chi)V +A = 0.$$
Solution of this equation that satisfy the initial conditions is equal to:
\begin{equation}\label{sol2}
V(\chi,\tau) = \frac{\lambda+i\chi-\sqrt{(\lambda+i\chi)^2 -4A\tau}}{2\tau}
\end{equation}
The branch of the square root in (\ref{sol2}) is chosen to provide zero
asymptotic for $V$ at infinity.

General solution of (\ref{quadZ}) with the velocity $V(\chi,t)$ satisfying
the Hopf's equation is given by the formula
$$
z(\chi,\tau) = G(i\chi-\tau V),
$$
with arbitrary function $G$. From the initial conditions (\ref{IVP}) one
can easily obtain that
$$G(\xi) = -i\xi,$$
and for $z(\chi,\tau)$ we get the expression:
\begin{equation}\label{sol1}
z(\chi,\tau) = -\frac{i}{2}\{-\lambda +i\chi +\sqrt{(\lambda+i\chi)^2
-4A\tau}\}.
\end{equation}
For $R(\chi,\tau)$ one can get the following formula:
\begin{equation}\label{R2}
R(\chi,\tau) = \frac{2\sqrt{(\lambda+i\chi)^2 -4A\tau}}
{\lambda+i\chi+\sqrt{(\lambda+i\chi)^2 -4A\tau}}.
\end{equation}

Let us consider behavior of $V(w,t)$ and $R(w,t)$ just after the cut
emerges, namely for
$$t << \frac{\lambda^2}{A}.$$

$R(w,t)$ has two branch points
\begin{eqnarray}\label{bra}
iw_1 &=& -\lambda -\int_0^\tau\frac{V_c}{R_c}d\tau +2\sqrt{A\tau},\cr
iw_2 &=& -\lambda -\int_0^\tau\frac{V_c}{R_c}d\tau -2\sqrt{A\tau}
\end{eqnarray}
Let us make an approximation to $V_c$ and $R_c$. We will estimate value
of $V$ and $R$ at
the point which is complex conjugate to the branch point $w_1$:
\begin{eqnarray}\nonumber
V_c &=& \mbox{c.c.}\left(V|_{iw = -iw_1}\right)
\simeq {1\over 2}\frac{A}{\lambda},\cr
\cr
R_c &=& \mbox{c.c.}\left(R|_{iw = -iw_1}\right)\simeq 1.
\end{eqnarray}

Branch point hits the real axes when $w_1$ becomes zero, at
\begin{equation}\label{tau0}
\tau = \tau_0 \simeq 0.34\frac{\lambda^2}{A}.
\end{equation}
In the vicinity of $w_1$, $R(w,\tau)$ behaves like this:
\begin{equation}\label{asym}
R(w,\tau) \simeq \frac{2}{(A\tau)^{\frac{1}{4}}}\sqrt{i(w-w_1)}
\end{equation}

(The latter must be deleted)

Now we can estimate $R_c(t)$. According to (\ref{apprU}) and (\ref{apprB})
it is equal to the value of the complex
conjugate function $\bar R$ at the point $w \simeq w_1$. Or, it is equal to
the value of $R$ at the point $w=-w_1$. Thus,
\begin{equation}\label{Rc}
R_c(\tau) = \frac{2\sqrt{1-\frac{2\sqrt{A\tau}}{\lambda}}}
{1-\frac{\sqrt{A\tau}}{\lambda}+\sqrt{1-\frac{2\sqrt{A\tau}}{\lambda}}}
\end{equation}
Recall that ``true'' time $t$ is related to $\tau$ through the
equation (\ref{newtime}) we can write down differential equation for time:
\begin{equation}\label{toftau}
dt = \frac{d\tau}{R_c(\tau)}.
\end{equation}
After integration of (\ref{toftau}) one can get
\begin{equation}\label{toftau2}
t = \frac{\tau}{2}+\frac{\lambda^2}{4 A}
\lbrace\frac{4}{5} +
\frac{1}{5}\left(1-\frac{2\sqrt{A\tau}}{\lambda}\right)^{5\over 2}
-\left(1-\frac{2\sqrt{A\tau}}{\lambda}\right)^{1\over 2}
\rbrace.
\end{equation}
For the small $t$ and $\tau$ relation (\ref{toftau2}) gives
$$
t \simeq \tau + \frac{A}{8\lambda^2}\tau^2.
$$
Rescaled time $\tau$ makes sense only for
$$
\tau \le \tau_0 = \frac{\lambda^2}{4A}.
$$
At $\tau = \tau_0$ ``true'' time tends to some finite value $t_0$:
$$
t \rightarrow t_0 = \frac{13}{10}\frac{\lambda^2}{4A}
$$
and in the vicinity of $\tau_0$ the following relation takes place:
$$
(\tau_0 -\tau) \simeq \frac{8A}{\lambda^2}(t_0 - t)^2.
$$
(The former must be deleted)

Of course, the 'approximation of narrow cut' is valid only for the time
$$
\tau << \tau_0,
$$
but formally it can be extended to $\tau_0$, and one can obtain the
limiting value of $R(w,\tau_0)$ from (\ref{R2}) or (\ref{asym}):
\begin{equation}\nonumber
R(w,\tau_0) \simeq \sqrt{\frac{1}{\lambda}}\sqrt{iw}.
\end{equation}
At the surface ($w=u$)
\begin{equation}\label{umber}
R(w,\tau_0) \simeq \sqrt{\frac{|u|}{\lambda}}(1+i\mbox{sign}(u)).
\end{equation}
The slope of the surface is given by the ratio:
$$
\frac{\partial y}{\partial x} = -\frac{\mbox{Im}\{R\}}{\mbox{Re}\{R\}} =
-\mbox{sign}(u).
$$
It means that the $90^o$ angle appears at the surface. Again, this
angle-type singularity takes place only for approximate equations
(\ref{quadZ}) and (\ref{quadV2}).

\section{Dynamics of zeroes and poles}
In this section we consider evolution of another initial condition for
$R$, namely, instead of initially flat surface we will consider
\begin{eqnarray}\label{iniR}
R(w,0) = 1+ \frac{a}{b+iw},
\end{eqnarray}
and the same (\ref{IVP}) for $V$.
$R(w,0)$ has simple pole at $w=ib$ and simple zero at $w=i(a+b)$.
Also it has right asymptotic behavior at infinity. Here we make important
suggestion for the latter: pole of $R$ is far from the cut of $V$
It is obvious that this pole does not produce new singularity in $V$,
and the
approximations (\ref{apprU}) and (\ref{apprB}) hold ( at least for some
time). The only difference is that for $U$ additional term appears in
(\ref{apprU}), namely:
$$U \rightarrow U + \frac{\mbox{const}}{b+i\chi-\tau V}.$$

\section{Self-similar solutions}
Solution (\ref{sol2}) and (\ref{R2}) are self-similar for the variable:
\begin{equation}\label{selsim}
\xi = \frac{\lambda +i\chi}{\tau}.
\end{equation}

\newpage

\appendix
\section{Poisson's Bracket}\label{ap:PoissonBraket}

Let us consider Poisson bracket between Hamiltonian $\cal H$ and some functional
$$\alpha(z,\bar z,\Phi,\bar\Phi).$$
\begin{equation}\label{PB}
\{\alpha,{\cal H}\} = \int\left (\frac{\delta\alpha}{\delta z}\dot z
+\frac{\delta\alpha}{\delta{\bar z}}\dot{\bar z}
+\frac{\delta\alpha}{\delta\Phi}\dot\Phi
+\frac{\delta\alpha}{\delta{\bar\Phi}}\dot{\bar\Phi}
\right)du
\end{equation}
Every part of the above integral we calculate separately.
$$\int\frac{\delta\alpha}{\delta z}\dot z du =
4i\int\frac{1}{J}\hat P^+\left\{z^\prime\frac{\delta\alpha}{\delta z}\right\}
\left( \hat P^-\frac{\delta \cal H}{\delta\bar\Phi}
+\hat P^+\frac{\delta \cal H}{\delta\Phi}\right)du$$
$$\int\frac{\delta\alpha}{\delta\bar z}\dot{\bar z} du =
-4i\int\frac{1}{J}\hat P^-\left\{\bar z^\prime
\frac{\delta\alpha}{\delta\bar z}\right\}
\left( \hat P^-\frac{\delta \cal H}{\delta\bar\Phi}
+\hat P^+\frac{\delta \cal H}{\delta\Phi}\right)du$$
\begin{eqnarray}\nonumber
\int\frac{\delta\alpha}{\delta\Phi}\dot\Phi du &=&
4i\int\frac{1}{J}\hat P^+\left\{
\Phi^\prime\frac{\delta\alpha}{\delta\Phi}\right\}
\left( \hat P^-\frac{\delta \cal H}{\delta\bar\Phi}
+\hat P^+\frac{\delta \cal H}{\delta\Phi}\right)du+\cr
&+&4i\int\frac{1}{J}\hat P^+\left\{
\frac{\delta\alpha}{\delta\Phi}\right\}
\left(
\hat P^-\left\{\bar\Phi^\prime\frac{\delta \cal H}{\delta\bar\Phi}\right\}
-\hat P^+\left\{\Phi^\prime\frac{\delta \cal H}{\delta\Phi}\right\}
\right)du\cr
&+&4i\int\frac{1}{J}\hat P^+\left\{
\frac{\delta\alpha}{\delta\Phi}\right\}
\left(
\hat P^-\left\{\bar z^\prime\frac{\delta \cal H}{\delta\bar z}\right\}
-\hat P^+\left\{z^\prime\frac{\delta \cal H}{\delta z}\right\}
\right)du
\end{eqnarray}
\begin{eqnarray}\nonumber
\int\frac{\delta\alpha}{\delta\bar\Phi}\dot{\bar\Phi} du =
&-&4i\int\frac{1}{J}\hat P^-\left\{
\bar\Phi^\prime\frac{\delta\alpha}{\delta\bar\Phi}\right\}
\left( \hat P^-\frac{\delta \cal H}{\delta\bar\Phi}
+\hat P^+\frac{\delta \cal H}{\delta\Phi}\right)du+\cr
&+&4i\int\frac{1}{J}\hat P^-\left\{
\frac{\delta\alpha}{\delta\bar\Phi}\right\}
\left(
\hat P^-\left\{\bar\Phi^\prime\frac{\delta \cal H}{\delta\bar\Phi}\right\}
-\hat P^+\left\{\Phi^\prime\frac{\delta \cal H}{\delta\Phi}\right\}
\right)du\cr
&+&4i\int\frac{1}{J}\hat P^-\left\{
\frac{\delta\alpha}{\delta\bar\Phi}\right\}
\left(
\hat P^-\left\{\bar z^\prime\frac{\delta \cal H}{\delta\bar z}\right\}
-\hat P^+\left\{z^\prime\frac{\delta z}{\delta\Phi}\right\}
\right)du.
\end{eqnarray}
Finally, the Poisson bracket reads:
\begin{eqnarray}\label{PB1}
\left\{\alpha,{\cal H}\right\} &=&
4i\int
\frac{1}{J}\left(\hat P^+\left\{
\frac{\delta\alpha}{\delta\Phi}\right\}
+\hat P^-\left\{\frac{\delta\alpha}{\delta\bar\Phi}\right\}\right)
\left( \hat P^-\left\{\bar z^\prime\frac{\delta \cal H}{\delta\bar z}\right\}
-\hat P^+\left\{z^\prime\frac{\delta \cal H}{\delta z}\right\}\right)du+\cr
&+&4i\int\frac{1}{J}\left( \hat P^+\left\{
\frac{\delta\alpha}{\delta\Phi}\right\}+
\hat P^-\left\{\frac{\delta\alpha}{\delta\bar\Phi}\right\} \right)
\left(
\hat P^-\left\{\bar\Phi^\prime\frac{\delta \cal H}{\delta\bar\Phi}\right\}
-\hat P^+\left\{\Phi^\prime\frac{\delta \cal H}{\delta\Phi}\right\}
\right)du+ \cr
&+&4i\int
\frac{1}{J}\left(\hat P^+\left\{z^\prime\frac{\delta\alpha}{\delta z}\right\}
-\hat P^-\left\{\bar z^\prime\frac{\delta\alpha}{\delta\bar z}\right\}\right)
\left( \hat P^-\frac{\delta \cal H}{\delta\bar\Phi}
+\hat P^+\frac{\delta \cal H}{\delta\Phi}\right)du+  \cr
&+&4i\int
\frac{1}{J}\left(\hat P^+\left\{\Phi^\prime\frac{\delta\alpha}{\delta\Phi}\right\}
-\hat P^-\left\{\bar\Phi^\prime\frac{\delta\alpha}{\delta\bar\Phi}\right\}\right)
\left( \hat P^-\frac{\delta \cal H}{\delta\bar\Phi}
+\hat P^+\frac{\delta \cal H}{\delta\Phi}\right)du
\end{eqnarray}
Easy to see that $\left\{\cal H,\cal H\right\}\equiv 0.$

\section{On the cubic equations}\label{Appendix2}

\section{Hydrodynamics in the electric and magnetic field}
To be written.

\section{Useful formulas}
\begin{description}
\item
$$y^\prime = \hat Hx^\prime$$
\item
$$(x^\prime +\hat Hy^\prime \cdot )(x^\prime -\hat Hy^\prime \cdot ) f = J f$$
\item
$$\hat H(A\cdot B) = \hat H(\hat HA\cdot\hat HB) + A\hat HB +B\hat H A$$
\item
$$\hat H\frac{x^\prime}{J} = -\hat H\frac{y^\prime}{J}$$
\item{$\bullet$}
$$P.V.\int_a^b \frac{\sqrt{(s'-a)(b-s')}}{s'-s}ds' = \pi (\frac{a+b}{2}-s)$$
\item{$\bullet$}
$$\int_a^b \frac{\sqrt{(s'-a)(b-s')}}{s'+s}ds' =
\pi (\frac{a+b}{2}+s-\sqrt{(b+s)(a+s)})$$
\item{$\bullet$}
$$A^-B^- - A^+B^+ = {1\over 2}[
(A^- -A^+)(B^- +B^+)+(A^- +A^+)(B^- -B^+)].$$
\end{description}


\begin{thebibliography}{}

\bibitem[Dyachenko (2001)]{D2001}
\textsc{Dyachenko, A. I.} 2001
{On the dynamics of an ideal fluid with a free surface.}
\textit{Doklady Mathematics} \textbf{63}, 115--118.

\bibitem[Dyachenko, Kuznetsov, Spector \& Zakharov (1996)]{DKSZ96}
\textsc{Dyachenko, A. I., Kuznetsov, E. A., Spector, M.D. \&
Zakharov, V. E.} 1996
{Analytical description of the free surface dynamics of an ideal fluid
(canonical formalism and conformal mapping).}
\textit{Phys.~Lett.~A} \textbf{221}, 73--79.

\bibitem[Dyachenko, Kuznetsov \& Zakharov]{DKZ96}
\textsc{Dyachenko, A. I., Kuznetsov, E. A. \& Zakharov, V. E.} 1996
{Nonlinear dynamics of the free surface of an ideal fluid.}
\textit{Plasma Physics Reports} \textbf{22}, 829--840.

\bibitem[Dyachenko \& Zakharov (1996)]{DZ96}
\textsc{Dyachenko, A. I. \& Zakharov, V. E.} 1996
{Toward an integrable model of deep water.}
\textit{Phys.~Lett.~A} \textbf{221}, 80--84.

\bibitem[Galin (1945)]{G45}
\textsc{Galin, L. A.} 1945
{Unsteady filtration with a free surface.}
\textit{Comptes Rendus (Doklady) de L'Academie des Sciences de l'URSS}
\textbf{47}, 246--249.

\bibitem[Kuznetsov, Spector \& Zakharov (1993)]{KSZ93}
\textsc{Kuznetsov, E.A., Spector, M. D. \& Zakharov, V. E.} 1993
{Surface singularities of ideal fluid.}
\textit{Phys.~Lett.~A} \textbf{182}, 387--393.

\bibitem[Longuet-Higgins (1976)]{L-H76}
\textsc{Longuet-Higgins, M. S.} 1976
{Self-similar, time-dependent flows with a free surface.}
\textit{J.~Fluid Mech.} \textbf{73}, 603--620.

\bibitem[Polubarinova-Kochina (1945)]{PK45}
\textsc{Polubarinova-Kochina, P. Ya.} 1945
{On the displacement of the oil-bearing contour.}
\textit{Comptes Rendus (Doklady) de L'Academie des Sciences de l'URSS}
\textbf{47}, 250--254.

\bibitem[Zakharov (1968)]{Z68}
\textsc{Zakharov, V. E.} 1968
{Stability of periodic waves of finite amplitude on the surface
of a deep fluid.}
\textit{J.~Appl.~Mech.~Tech.~Phys.} \textbf{9}, 190--194.

\bibitem[Zakharov \& Dyachenko (1996)]{ZD96PD}
\textsc{Zakharov, V. E. \& Dyachenko, A.I.} 1996
{High-Jacobian approximation in the free surface dynamics of an ideal fluid.}
\textit{Physica D} \textbf{98}, 652--664.


\bibitem[Zubarev (2000)]{Zub2000}
\textsc{Zubarev, N. M.} 2000
{Charged-surface instability development in liquid helium: an exact solution.}
\textit{JETP Letters} \textbf{71}, 367--369.

\end{thebibliography}
\end{document}